# Switchable out-of-plane polarization in two-dimensional LiAlTe$_2$

Zhun Liu[1], Yuanhui Sun[1], David J. Singh[2,*], and Lijun Zhang[1,*]

[1]*Key Laboratory of Automobile Materials of MOE and School of Materials Science and Engineering, Jilin University, Changchun 130012, China.*

[2]*Department of Physics and Astronomy, University of Missouri, Columbia, MO 65211-7010 USA*

*Author to whom correspondence should be addressed: singhdj@missouri.edu, lijun_zhang@jlu.edu.cn

**Abstract:**

Covalent-polar semiconductors that show intrinsic two-dimensional (2D) vertical polarization present new device opportunities. These materials differ from ordinary ferroelectrics in that they are able to maintain polarization normal to a surface even with an unscreened depolarization field. Identifying phases that exhibit intrinsic 2D vertical polarization is an ongoing challenge. Here we report via computational material design the discovery of a new promising phase, specifically 2D LiAlTe$_2$. The design idea is developed from the physical understanding of three-dimensional hyperferroelectric covalent polar semiconductors. We used the structure determination method combining swarm intelligence algorithm and first-principles calculations to identify energetically stable structures. In addition to the expected layered version of bulk LiAlTe$_2$, β-LiAlTe$_2$, we find a novel 2D structure, γ-LiAlTe$_2$. In this phase, the vertical dipole can be switched between 0.07 and -0.11 e·Å. This switching is triggered by the movement of Li atom between two local energy minima. The associated asymmetric double-well energy profile can be continuously tuned by the applied electric field as well as strain. There is, therefore, a reversible transition between two polar states. This discovered off-plane switchability provides an opportunity for the 2D γ-LiAlTe$_2$ based interfacial phase change memory device for example by growing γ-LiAlTe$_2$/GeTe heterostructures.

## 1. Introduction

Controlling electric dipoles in the ultrathin films is a key challenge for due to the finite screening length of surface charges.[1] This is a consequence of the depolarizing field, which becomes more important as the ferroelectric film thickness is reduced. The lower limit for ferroelectric is controlled by the electrodes, in particular the screening length.[1b] However, recently, a different class of materials, specifically covalent polar semiconductors, also called hyperferroelectrics, have been identified.[2] These have stable polar states in ultrathin atomic layers and do require the screening of depolarizing fields, as was noted also in certain improper ferroelectrics.[3] This is important for the electronic polarizability that plays an important role in this class of materials, and underlies the hyperferroelectric behavior, specifically the resistance to depolarizing field. The key condition is the instability of longitudinal optic modes in addition to active transverse optic instability characteristic of normal ferroelectrics. This provides a mechanism for canceling the depolarizing field. As shown in **Figure 1**, a number of these hyperferroelectrics adopt a polar stuffed wurtzite structure which is typically formed by an ABC stoichiometry (A, B, and C represent different elements).[4] This mechanism motivates us to extend the range of materials by searching for a different class of covalent-polar compounds, $ABC_2$, exemplified by $LiAlTe_2$.[5] It provides opportunities for designing switchable devices since it is compatible with substrates such as Germanium Tellurium (GeTe), which are prototypical phase change memory materials.

The $ABC_2$ ternary compound $\beta$-$LiAlTe_2$ has a layered polar covalent tetrahedral structure, space group, P3m1 (156).[5a] As shown in Figure 1(a), this quadruple-layered (QL) crystal is formed by an in-plane covalent bonded $AlTe_4$ tetrahedral layer with Li atoms located in the distorted trigonal antiprisms, with van der Waals (vdW) bonding between the QL. This vdW bonding suggests that in addition to being polar, such material may be an amenable formation in single QL form as a 2D material. This raises the possibility of having a hyperferroelectric 2D layered material based on the phase with an out-of-plane polarization. In most case, there are two major approaches to obtain polarization in 2D materials: (a) functionalization of non-polar 2D materials; (b) searching for 2D intrinsic ferroelectric materials.[6] The former strategy uses chemical doping or surface passivation to produce switchable dipoles.[7] The latter approach is to find materials with intrinsic 2D polarization, such as Janus 2D materials.[8] These derivative 2D materials with broken mirror symmetry attracted considerable attention because of their unique properties, such as large piezoelectric effect, Rashba spin splitting, and second-harmonic generation response.[9]

In this work, we systematically explore the 2D vertical polarization in a single $LiAlTe_2$ QL has by studying the polar layers in the vdW type β-$LiAlTe_2$. We have discovered a series of metastable polar tetrahedral structures by global optimization structure search. In addition to the expected layered version of the bulk structure β-$LiAlTe_2$, we find for the first time that a novel 2D structure, γ-$LiAlTe_2$. We further investigated the low energy metastable structures and find switchable behavior based on distortions describable as primarily Li ion displacements. For the γ-(abba) stacking, the vertical dipole moment can be switched between 0.07 and -0.11 e·Å as the Li atom is shifted up or down between the two local minima. The electric field across the QL can induce a potential shift between the two surfaces, thus directly leading to the opposite electronic band alignments with respect to the vacuum level thereby changing the band gaps, e.g., the direct bandgap in the γ-up structure (1.57 eV) is 0.26 eV larger than in the γ-down structure (1.31 eV). The asymmetric double-well energy profile can be continually tuned by the applied electric and strain. Consequently, a twin double-well potential with moderate transition barrier can be obtained (~75 meV for -5% strain). Based on these results we propose 2D γ-$LiAlTe_2$/GeTe vdW heterostructures for interfacial phase change memory due to their compatible lattices as well as switchable interlayer electronic coupling (metal-semiconductor switching).

## 2. Computational Methods

First-principles calculations are carried out by using the plane-wave pseudopotential approach within density functional theory (DFT) as implemented in the Vienna Ab Initio Simulation Package (VASP).[10] The electron-core interactions are described by the projected augmented wave pseudopotentials and the exchange and correlation functional was treated using the Perdew-Burke-Ernzerhof (PBE) parametrization of GGA for structural relaxations and total energy calculation.[11] To better describe the interlayer interactions, we use the optB86b vdW-DF implemented in VASP to treat dispersion interactions.[12] The optimized the lattices for the bulk phase β-LiAlTe$_2$ (space group: P3m1) is 4.33 (Å) and 7.20 (Å) for the *a* and c axis, which is slightly contracted by 3% and elongated by 1.5% with respect to the experiment parameters, respectively. We used global optimization to identify stable or metastable phases for single-layered structure by the CALYPSO (Crystal structure analysis by particle swarm optimization) software package.[13] We used hybrid functional of Hyed-Scuseria-Ernzerhof (HSE06) for band structure calculations. The electronic minimization was performed with a tolerance of $10^{-6}$ eV, and ionic relaxation was performed with a force tolerance of 0.005 eV Å$^{-1}$ on each ion. The kinetic energy cutoff was set as 520 eV, and the k-points mesh was used with a grid spacing of less than 2π×0.15Å$^{-1}$. The climbing image nudged elastic band method is used to determine the energy barriers of the kinetic phase change process. The phonon spectra were calculated using the finite displacement method. The interatomic forces were computed with a 5×5×1 supercell and a 7×7×1 k-point mesh using the VASP code. The tolerance for the force convergence used for the phonon calculations was $10^{-8}$ eVÅ$^{-1}$. The proposed 2D γ-LiAlTe$_2$/GeTe vdW heterostructures were optimized with a commensurate alignment of the materials.

## 3. Results and Discussion

### 3.1. Polarization properties of bulk LiAlTe$_2$

We start from investigating electric dipole polarization properties of bulk LiAlTe$_2$, β-LiAlTe$_2$. The structure is polar as expected from a stacking ABC along the hexagonal c-axis, which necessarily breaks inversion symmetry.[6a] Our calculated vertical spontaneous polarization (electron + ionic contribution) is ~37 μC/cm$^2$ as obtained using the Berry phase approach. This robust vertical polarization is comparable to the bulk ferroelectric in PbTiO$_3$ (~50 μC/cm$^2$).[1b] To access the switchability for β-LiAlTe$_2$, we calculated the kinetic barrier for polarization reversal through the centrosymmetric reference state. The resulting barrier is 0.14 eV/atom, which is too high for realistic switchable device applications. This is due to the covalent bond breaking needed to overcome the barrier.[4, 6a, 14]

### 3.2. Crystal structure search of 2D LiAlTe$_2$

The robust vertical polarization in vdW bonded β-LiAlTe$_2$ motivates the exploration of possible polar structures in two-dimensional (2D) counterparts with thickness down to a QL limit. In general, there are two major approaches to obtain polarization in 2D materials: (a) functionalization of non-polar 2D materials; (b) searching for 2D intrinsic ferroelectric materials.[6] The former strategy uses chemical doping or surface passivation to produce switchable dipoles.[7] The latter approach is to find materials with intrinsic 2D polarization.[8] For example, it was suggested based on experimental trends in the dielectric constant with thickness and first-principles calculations that single layer In$_2$Se$_3$ may exhibit room ferroelectricity.[15] One complication is that the structure of a single layer may differ from the layer structure in bulk. We used global structure optimization[13a] to identify the stable or metastable phases in an isolated 2D layered. As shown, in the right plane of Figure 1(b), the ground state 2D structure emerged in the third generation of the optimization and no lower energy structures were found from the continuation of the search.

As illustrated in Figure 1(b) (Left panel), the ground state structure (GS) has a triple-layered tetragonal configuration, in which the Al and Li atoms alternate in the middle atomic plane while the Te atoms are located on both sides. Thus the ground state maintains a nonpolar Tasker type II surface. The remaining low energy structures can be divided into three groups by the Al-Te structural frameworks. The first group of metastable structures has two basic types of structures with energy very close to the ground state (~ 0.05 eV/atom higher). The differences between the structures in the first group come from the Li atom locations. As highlighted by the red frames, the β-phase QL is reproduced with abbc stacking similar to the bulk phase, while γ-phase tetrahedral QL exhibits abba stacking and shifts the Li atom below the surface. To describe bonding sequence in the Li-Te-Al-Te QL, we denote them as γ-(abba) and β-(abbc), respectively (explicit structural information in Table S1 of Supporting Information). These cation ordered polymorphs with Tasker type III surfaces exhibit different out-plane dipole moments.[8f] Thus it would be possible in principle to design dipole electric devices that involve switching the structure between these polar states under different gated field.[1a, 6b, 16] The structures in the second group are about 0.13 eV/atom above the ground state energy. These show similar tetragonal frameworks to the ground state but with Li atoms separated from the Al atomic plane with different vertical distance. The third group in our list of structures show distorted octahedral coordination for the Al-Te framework and have energies least 0.15 eV/atom higher than the ground state.

We now focus on the polar structures in the first group. As highlighted by the red frameworks in Figure 1(b), the small energy differences between these structures can be understood from their similar structures. The difference between them comes from the Li atom positions. Transitions between them can be viewed as Li atom displacements. As illustrated in Figure 2(a), the optimized transition pathway from the β structure to the γ structure is hopping of the Li atom from the c site to the a site by jumping over a barrier. This hopping barrier is very high (0.64 eV/f.u.) for the Li atom on the ridge site due to the reduced coordination implying that this transition is infeasible as a switching mechanism.

The next question is whether the covalent bonding of the structure is sufficiently strong to maintain these polar structures including the effects of depolarizing fields. We choose a 2 × 1 orthogonal supercell to study the transformation of the γ structure to the ground state. As shown in Figure 2(b), the transition barrier is approximately 0.72 eV/f.u., suggesting the covalent bonding is sufficient to maintain this structure. We checked dynamic stability by calculation of the phonon spectra for the 2D hexagonal polar structures. As shown in Figure 2 (c-d), the harmonic phonon spectra at 0K for the β- and γ-structures are stable (note that the calculated flexural branch has a slight instability near the zone center, is below the precision of the calculation).

**3.3. Polar nature in two-dimensional hexagonal LiAlTe$_2$**

In order to gain insight into the vertical polarization correlated with the Li atom displacement in 2D hexagonal LiAlTe$_2$, we displace the Li with respect to the top surface and track the energy profile and the dipole. This involves constructing a series of structures with vertical shifts from the starting positions (position in the γ-(abba) or β-(abbc) structures). As shown by the blue curves in Figure 3(a), the asymmetric energy profiles for two positions show a markedly different character. There are two local energy minima for the γ-(abba) stacking but only one for the β-(abbc) stacking. We thus use the γ-(abba) structure as an example to compare the total energy for the double size structure (2×1×1 supercell) with different vertical polarization in the two cells. We confirm that the double size structure with opposite polarization can be relaxed to a steady configuration. As shown in Table S2, the Li atoms occupied with opposite polar sites (dn-up) is approximately 26 meV/f.u. lower than the Li atoms with both up polar states (up-up), but is still 13 meV/f.u. higher than the Li atoms with both down polar states (dn-dn). This also implies that the opposite polar states would reset to the lower energy down polar state if suitably perturbed. Interestingly, the residual dipole moment is around -0.02 e·Å when the Li atom is at the surface for both curves. As the Li atom moves the dipole changes. For the abba stacking, the vertical dipole moment is switched between 0.07 and -0.11 e·Å as the Li atom shifted up or down between the two local minima. The sign change of the dipole between the γ-up and γ-down states is also clearly seen in the opposite potential drops along the vertical direction in Figure 3 (b).

The different behaviors of the abba and abbc stackings can be understood in terms of the electron density. Figure 3 (c) displays the planar averaged charge density changes in the polar structures with Li atoms bonding at different sites. Since the abbc stacking does not have a local minimum, we construct a β-down structure with the same downward Li shift as in the γ-down structure. In the γ-up and β-up polar structures, there is a substantial electron depletion above the top surface and accumulation between on the Li-Te bonding centers. This is similar between the two structures. However, with Li atoms below the surface, the results are different. The γ-down structure shows an additional charge accumulation between the Li atom and the Te atom below it at the bottom surface. This is stabilizing and explains the deeper well.

Normal ferroelectrics exhibit two inverted polar ground states with completely equal electronic properties.[6a] However, the displacive phase transition induced polar states in the 2D LiAlTe$_2$ show different characters. Owing to the existence of opposite out of plane electric dipole, the electric field across the QL can induce a large potential shift between the two surfaces, thus directly leading to the opposite electronic band alignments with respect to the vacuum level of the surface.[7a, 17] This is seen by comparing the electronic band structure of the two abba stacking polar structures. As shown in Figure 3 (d), the direct bandgap (1.57 eV) in the γ-up polar structure is 0.26 eV larger than for the γ-down structure. This enlarged band gap confirms the opposite potential drop induced a shift of energy levels with respect to the atomic orbital states on the top surfaces.[7a] From the orbital-projected band structure, we can see that the nearly aligned energy levels come from the Te atoms on the top surface, i.e. the Te 5s lone pairs (about -10 eV) and the Li-Te partial ionic bonding (valence band edge states). In contrast, for the Al-Te covalent hybrid states, the energy levels in γ-up structure (cyan dots) are upshifted by 0.25 eV compared to the γ-down structure (blue dots). Inversion symmetry breaking perpendicular to the surface combined with spin orbit coupling (SOC) may break spin degeneracy as in the Rashba effect.[18] Such an effect has been reported in 2D geometries. It depends on the spin-orbit coupling (SOC) strength and the electric field perpendicular to the surface.[19] We compared the bands including the SOC effects for γ-up and γ-down structures. As shown in Figure S1, the SOC splits the degenerate valence band maxima (Γ-point) by about 0.6 eV for both structures, thereby reducing the bandgap by about 0.23 eV. It should be noted that the intrinsic spin-orbital splitting at the K-point is about 0.3 eV for both valence bandages, while the Rasha splitting around the Γ-point for the band edge is negligible.

**3.4. Strain and electric field tunable switching in 2D γ-LiAlTe$_2$**

Next, we examine the possibility of reversibly switching between the two minima, which as mentioned are non-degenerate. In particular the issue is whether it is possible to tune between these two minima with field and whether the critical field for this, i.e. the coercive field is of a realizable magnitude.[6a, 14] As shown in Figure 4 (a), the asymmetric Li potential energy profile shows that the γ-down polar state is preferred without an external electric field. However, when a negative electric field gradually increased to -0.48 eV/Å, the electric polarization energy stabilizes the γ-up state due to the interaction of the dipole with the field. We note that 0.48 eV/Å is a very large field.

An important feature of wurtzite polar materials is their great sensitivity to elastic stress.[6c, 20] The strain may then significantly modify the local bonding energy and thus the barriers and energy differences between the minima.[8e, 21] As shown in Figure 4 (c), these properties behave oppositely with compressive and tensile strain as expected. The tensile strain expands the in-plane Li-Te bonds, thus destabilizing local γ-up states and changing the asymmetric double-well energy surface into an asymmetric single well. While for compress strain, the reduced in-plane bond lengths, resulting in a higher transition barrier and more stabilization of the γ-up states. As a consequence, a near twin double-well potential with moderate transition barrier can be obtained (see Figure 4 (c-d), ~75 meV for -5% strain), which is a level suitable for switching.

**3.5. Polar switching for van der Waals interfacial phase-change memory**

The phase change induced polar switch in 2D γ-LiAlTe$_2$ provides the opportunity for engineering vdW devices. The large sensitivity to interlayer strain and electric coupling makes the choosing candidate materials crucial.[1a, 1b] Here, we propose 2D γ-LiAlTe$_2$ /GeTe vdW heterostructures for interfacial phase change memory due to their compatible lattices as well as switchable interlayer electronic coupling.[16a] The explicit structural information is shown in Table S1 of Supporting Information. As shown in Figure 5 (a), aligning the γ-LiAlTe$_2$ QL and the single GeTe layer maintains the double well energy profile. The transition barriers for switching two minima configurations are in the acceptable range of 80~150 meV along the vertical displacive pathway. We further confirm the stability and switchable behavior by first-principles molecular dynamic simulations at 300K (see Figure S2). More interesting, the electric coupling in γ-LiAlTe$_2$/GeTe vdW heterostructures exhibits markedly different behavior between two configurations. With the Li atom in up polar position, a lower surface work function in γ -LiAlTe$_2$ promotes electron transfer to GeTe, which results in spontaneous mutual doping.[7a] This can be clearly reflected by the conductive band edges from the projected band structures (electric doping in GeTe and hole doping in γ-LiAlTe$_2$), see in Figure 5 (b). On the contrary, the Li atom at the down polar state exhibits a slight electron accumulation between their vdW interfaces by electron transfer from both materials. As illustrated by the projected band structure at down polar state, the valence band edges are derived from Ge-Te and Li-Te orbitals, while the band gap remains moderate (0.75 eV). The different electronic coupling behaviors between the switched polar states (metal-semiconductor switching) imply that this type of vdW heterostructure may be promising for 2D interfacial phase change memory based on resistance readout and electric field switching.[16] It may also be possible to use this type of contrast to design other devices including ones that combine the polar switching with the phase change behavior of $Ge_2Sb_2Te_5$ (GST).[22] It will be of interest to determine whether these systems are experimentally realizable and to measure the resistive switching characteristics.

## 4. Conclusion

2D vertical polarization in a single $LiAlTe_2$ QL has been theoretically explored by studying the polar layers in the vdW type β-$LiAlTe_2$. Global optimization structure search yields a non-polar ground state for a single $LiAlTe_2$ QL. In addition, there exists a series of metastable polar tetrahedral structures. We investigated low energy metastable structures and find switchable behavior based on distortions describable as primarily Li ion displacements. For the γ-(abba) stacking, the vertical dipole moment can be switched between 0.07 and -0.11 e·Å as the Li atom is shifted up or down between the two local minima. The electric field across the QL can induce a potential shift between the two surfaces, thus directly leading to the opposite electronic band alignments with respect to the vacuum level thereby changing the band gaps, e.g., the direct bandgap in the γ-up structure (1.57 eV) is 0.26 eV larger than in the γ-down structure (1.31 eV). The asymmetric double-well energy profile can be continually tuned by the applied electric and strain. Consequently, a twin double-well potential with moderate transition barrier can be obtained (~75 meV for -5% strain). Based on these results we propose 2D γ-$LiAlTe_2$/GeTe vdW heterostructures for interfacial phase change memory due to their compatible lattices as well as switchable interlayer electronic coupling (metal-semiconductor switching).

**Supporting Information**
Supporting Information is available.

**Acknowledgements**
The work at Jilin University is supported by the National Natural Science Foundation of China (Grant No. 61722403 and 11674121) and Program for Jilin University Science and Technology Innovative Research Team. Work at the University of Missouri is supported by the U.S. Department of Energy, Basic Energy Science through Award Number DE- SC0019114.

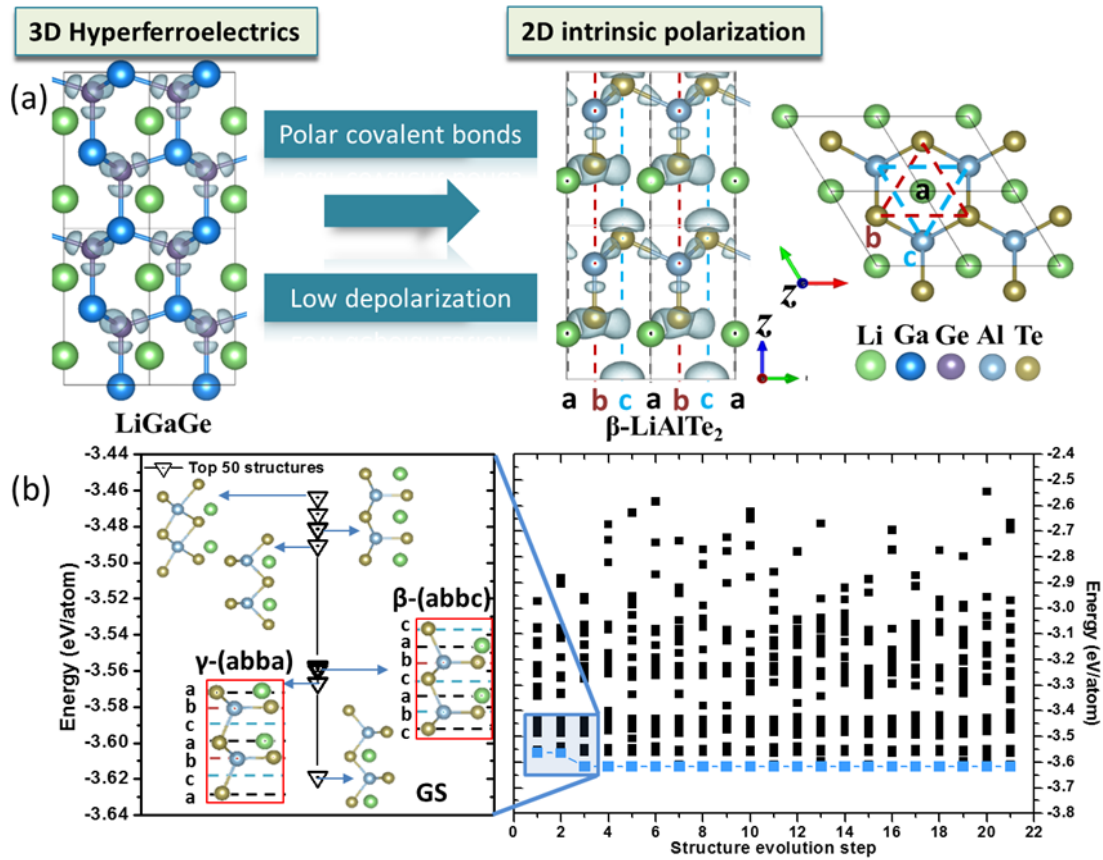

**Figure 1.** (a) Illustration of the role of polar covalent bonds against the depolarizing in steady hyperferroelectric semiconductors. Side and top view of the 2×2×2 structure model for bulk phase LiGaGe (Left panel) and layered β-LiAlTe$_2$ (Right panel). The electron localized function (ELF) with an isosurface of 0.8 is plotted in the side view. Considering the partial ionic character in Li atom, we only connect the Al-Te and Ga-Ge bonds to highlight the covalent tetrahedral structure character. The colored dashed lines in the figure denote three different atomic locations (a, b, c) in the hexagonal lattice. (b) Prediction of the 2D LiAlTe$_2$ structure based on particle swarm global optimization. Left panel: The insets show the representative 6 identified structures. The red frames highlight the polar structures with lower energy in the first group, which are denoted as γ-(abba) phase, and β-(abbc) phase by their atomic bonding character. Right panel: The evolution of the energy distributions during the structure search generations.

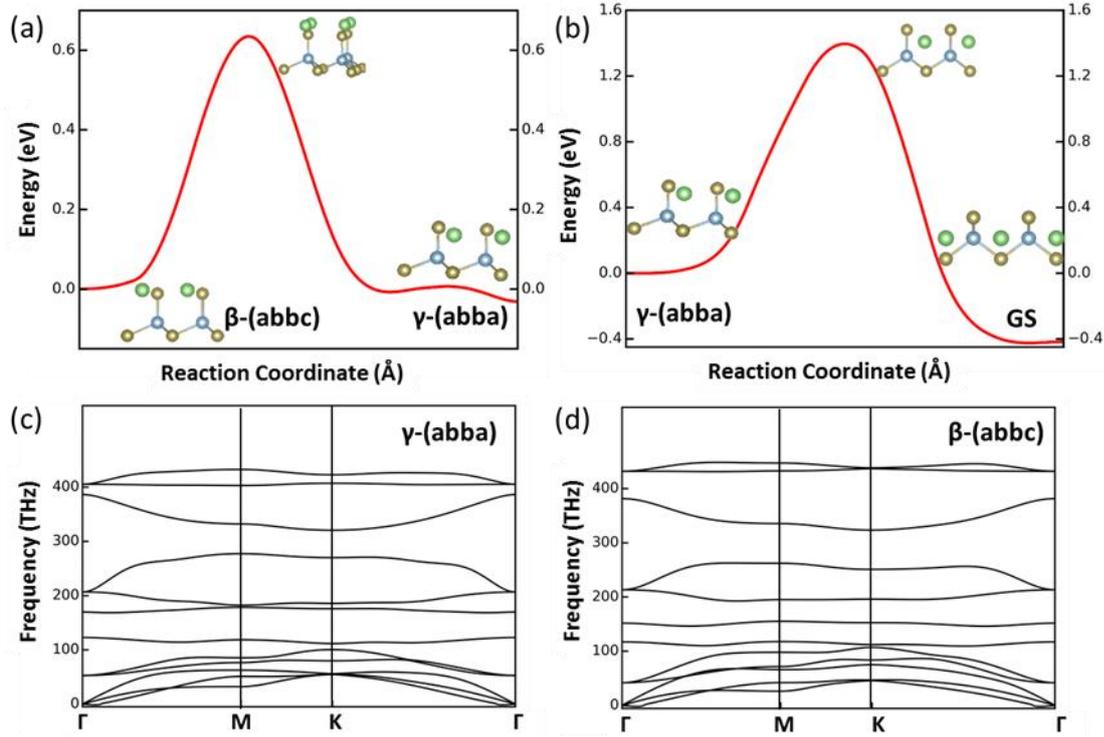

**Figure 2.** Kinetic pathways for phase transition between the two searched polar structures (from the β structure to the γ structure) is plotted in (a). (b) The energy profile for the represented 2D polar phase (γ) transforms into the nonpolar ground state (GS). Due to the different symmetries, we choose a $2 \times 1$ orthogonal supercell for the transformation of the structures. The insets indicate the initial, transition and final states to represent the atom motion during the phase transition. (c) and (d) show the calculated harmonic phonon spectra at 0K for the β and γ structures, respectively.

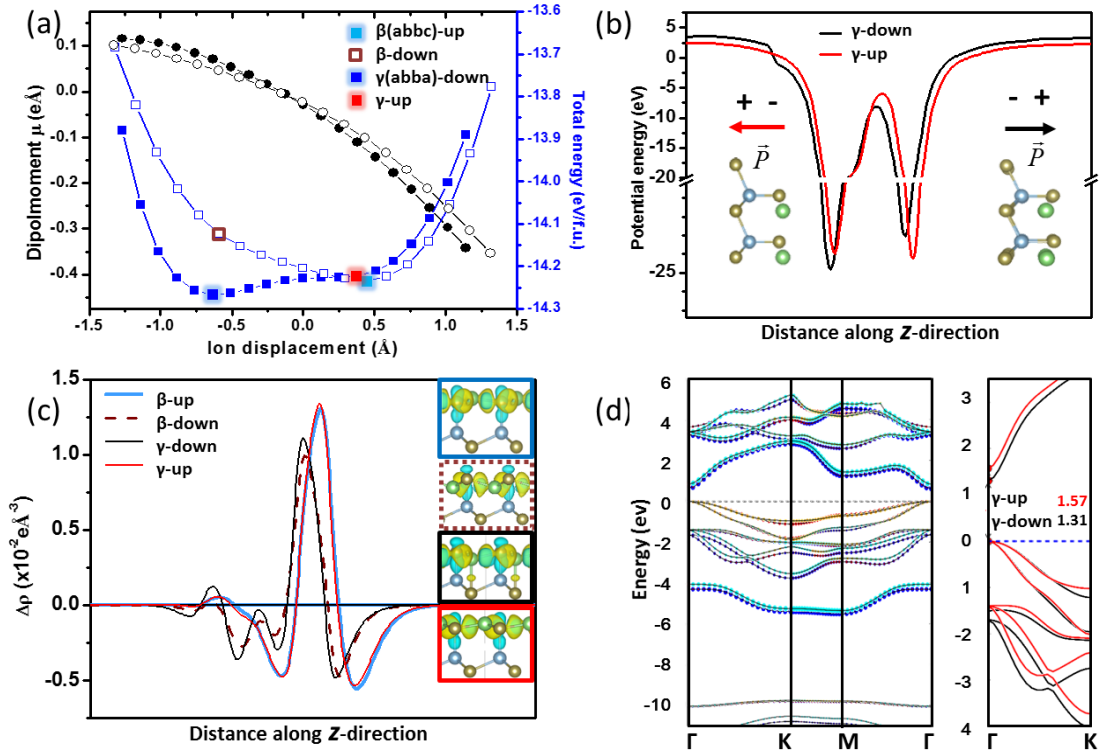

**Figure 3.** (a) Dipole and potential energy profile for the Li atoms displaced normal to the surface. Three colored points represent the structures with Li atom displacements at the local energy minimum. (b) The comparison of plane averaged electrostatic potential for abba stacking polar structures with opposite polarization. (c) The comparison of charge density difference for different 2D polar structures. The brown dashed line represents the charge density difference for a virtual β-down structure constructed by applying the same Li atom shift as in the γ-down structure. The insets display the corresponding charge density difference plotted with isosurfaces of $2 \times 10^{-3}$. (d) Comparison of band structures (left plane) and the HSE corrected band edges (right plane) for two abba stacking structures with opposite polarization. The cyan (blue) and yellow (red) dots have represented the bands contributed by the Al-Te and Li-Te orbital interactions in γ-up (γ-down) structures, respectively.

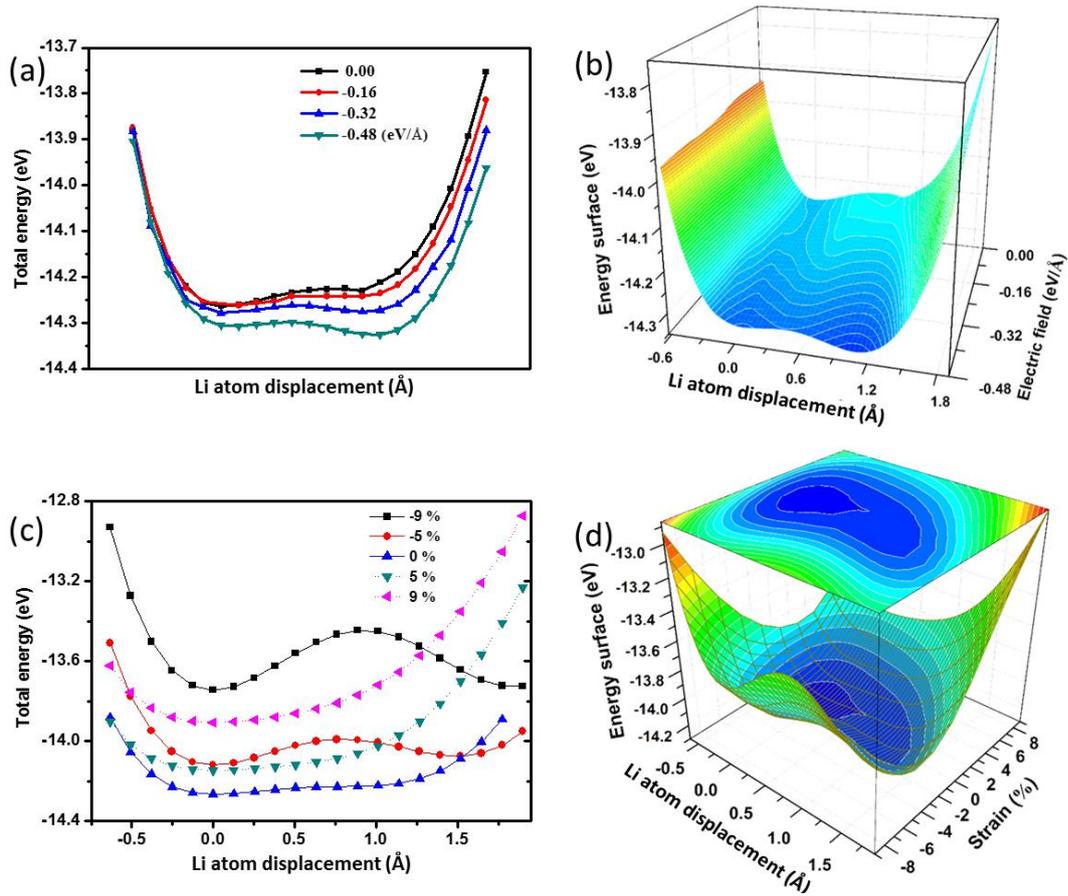

**Figure 4.** (a) Change of asymmetric double-well energy profile under the different magnitude of the electric field and the corresponding 3D displacive potential energy surface is plotted in (b). (c) Evolution of double-well energy profile with varied in-plane strain and its 3D potential energy surface is plotted in (d).

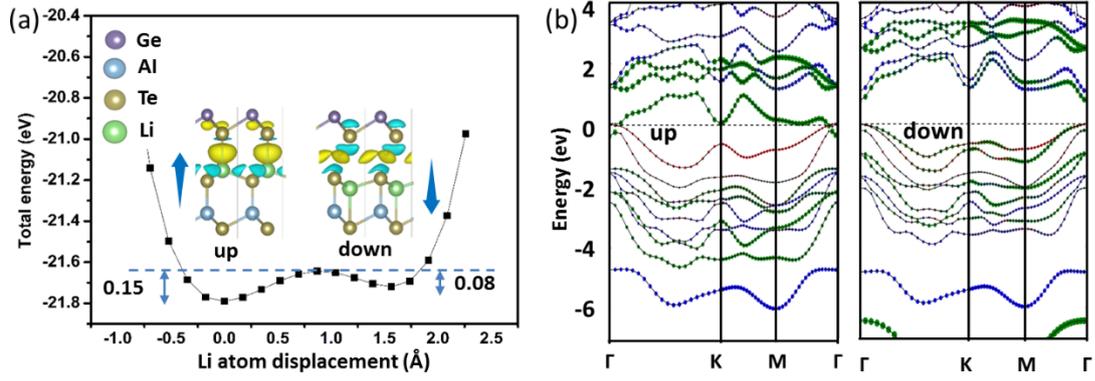

**Figure 5.** (a) Polar switch double-well potential for the 2D γ-LiAlTe$_2$/GeTe vdW heterostructures. The insets display the corresponding charge density difference plotted with isosurfaces of 8× 10$^{-3}$ and 2× 10$^{-3}$ for the up and down polar state in γ-LiAlTe$_2$, respectively. (b) Projected electronic band structures for two local minima polar states. Left (right): Li atom in up (down) displacement configuration. The green, red and blue dots are represented the electric bands contributed by the Ge-Te, Li-Te and Al-Te orbital interactions in the two polar structure, respectively.

A new two dimensional (2D) γ-LiAlTe$_2$ is discovered by first-principles calculations. The vertical dipole moment can be switched to opposite direction as the Li atom is shifted up or down between the two local minima. A 2D γ-LiAlTe$_2$/GeTe vdW heterostructures for interfacial phase change memory is further proposed due to their compatible lattices as well as off-plane metal-semiconductor switching.

**Keyword** two dimensional materials, switchable polarization, LiAlTe$_2$, Janus materials, interfacial memory

ToC figure

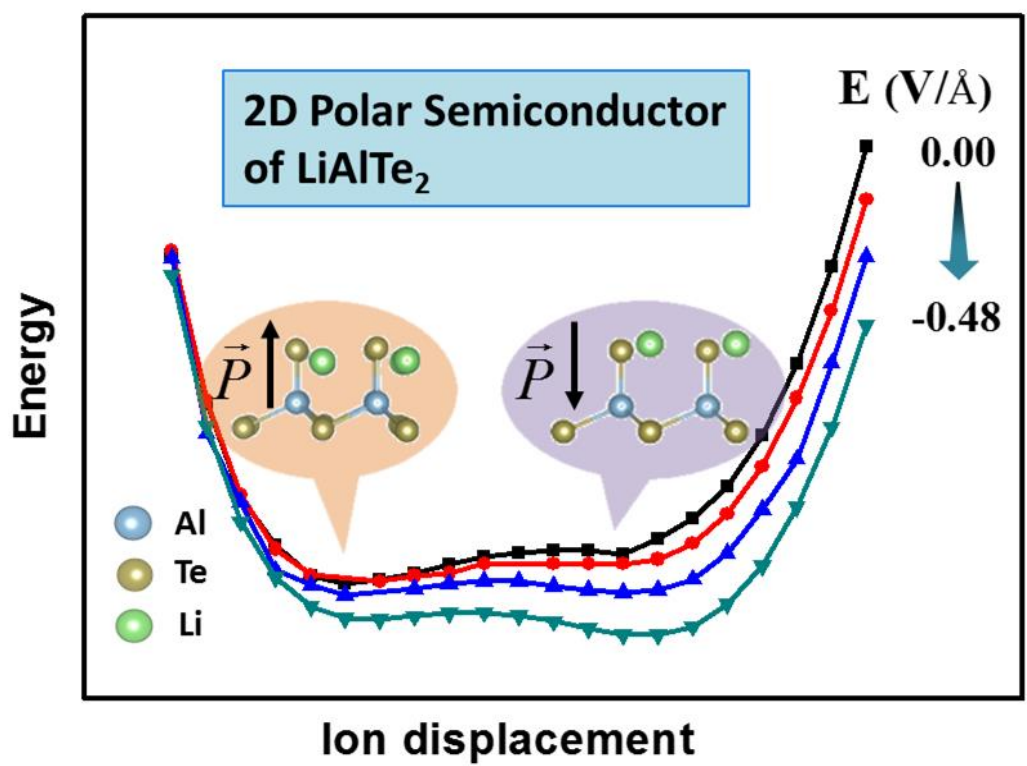

Supporting Information

# Switchable out-of-plane polarization in two-dimensional LiAlTe$_2$

Zhun Liu[1], Yuanhui Sun[1], David J. Singh[2,*], and Lijun Zhang[1,*]

[1]*Key Laboratory of Automobile Materials of MOE and School of Materials Science and Engineering, Jilin University, Changchun 130012, China.*
[2]*Department of Physics and Astronomy, University of Missouri, Columbia, MO 65211-7010 USA*
**Author to whom correspondence should be addressed*: singhdj@missouri.edu, lijun_zhang@jlu.edu.cn

**Table S1**. Structural data for the polar two-dimensional LiAlTe$_2$ identified from structure searches and the optimized γ-LiAlTe$_2$/GeTe heterostructures.

| Material Space group | Lattice parameters (Å) | Wyckoff positions | Atoms | x | y | z |
|---|---|---|---|---|---|---|
| γ-LiAlTe$_2$ (*P3m1*) | a= 4.40100 b= 4.40100 c=21.12230 | 1b | Li1 | 0.33333 | 0.66667 | 0.55786 |
| | | 1b | Te2 | 0.33333 | 0.66667 | 0.41064 |
| | | 1c | Te1 | 0.66667 | 0.33333 | 0.5867 |
| | | 1c | Al1 | 0.66667 | 0.33333 | 0.46708 |
| β-LiAlTe$_2$ (*P3m1*) | a= 4.39270 b= 4.39270 c=24.42700 | 1a | Li1 | 0 | 0 | 0.58198 |
| | | 1c | Te1 | 0.66667 | 0.33333 | 0.56522 |
| | | 1c | Al1 | 0.66667 | 0.33333 | 0.46111 |
| | | 1b | Te2 | 0.33333 | 0.66667 | 0.41397 |
| γ-LiAlTe$_2$ /GeTe (*P3m1*) | a=4.19260 b=4.19260 c=30.57050 | 1b | Li1 | 0.33333 | -0.33333 | 0.46663 |
| | | 1b | Te2 | 0.33333 | -0.33333 | 0.37245 |
| | | 1b | Te3 | 0.33333 | -0.33333 | 0.61741 |
| | | 1c | Ge1 | 0.66667 | -0.66667 | 0.32434 |
| | | 1c | Al1 | 0.66667 | -0.66667 | 0.57731 |
| | | 1c | Te1 | 0.66667 | -0.66667 | 0.49457 |

**Table S2.** The comparison of total energy for the double size γ-(abba) structure with the different vertical polar configuration in two cells.

| Polar structures | ev/f.u. |
|---|---|
| γ (dn-dn) | -14.265 |
| γ (dn-up) | -14.252 |
| γ (up-up) | -14.226 |

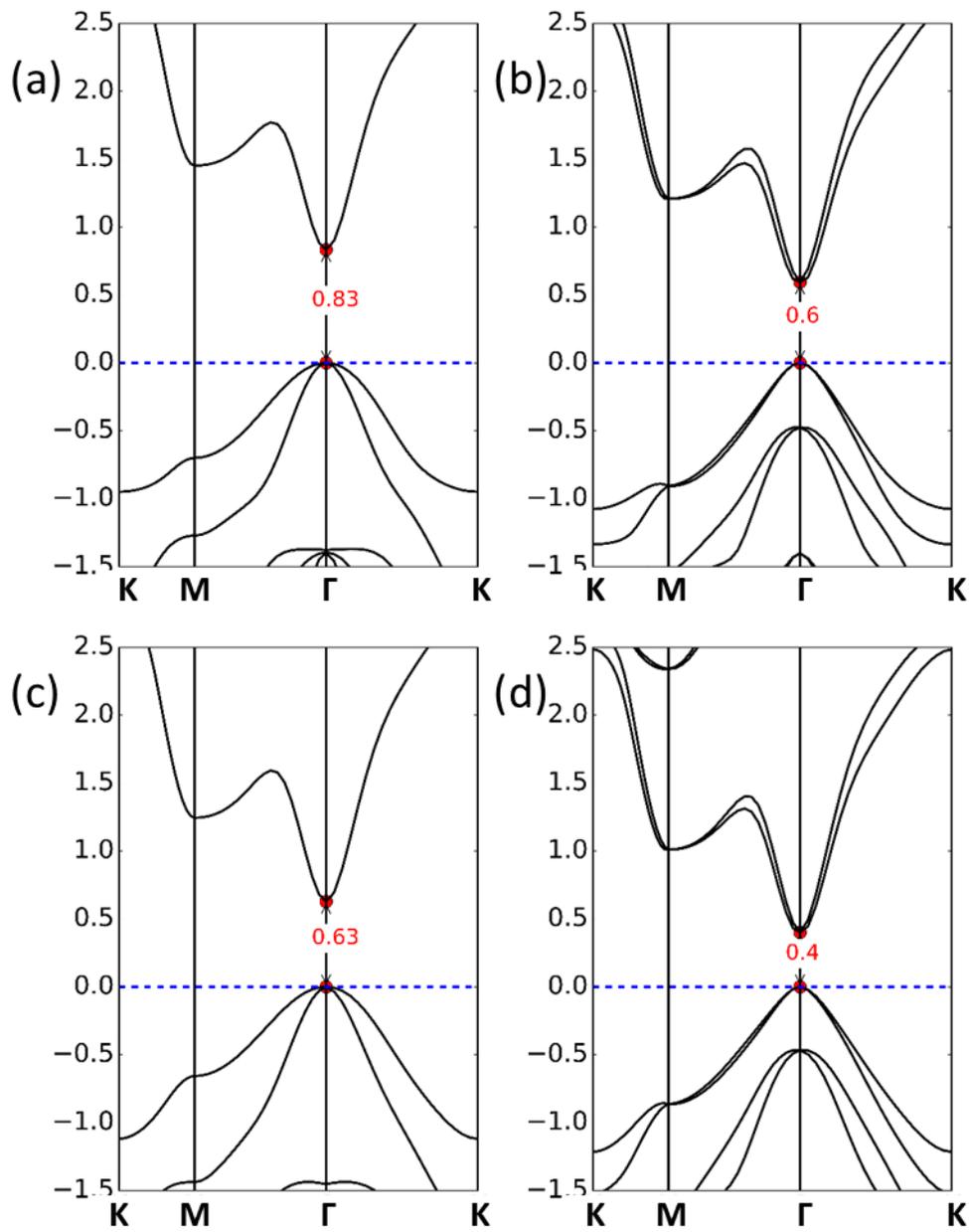

**Figure S1.** Electronic bands without (a) and with (b) SOC effects for 2D LiAlTe$_2$ $\gamma$-up structures and for $\gamma$-down in (c) and (d), respectively.

We utilized a 5×5×1 supercell for first-principles molecular dynamics (MD) simulations. We did simulations at 300 K for 6 ps with a time step of 1 fs. We observed displacements for Li atoms without additional structural reconstruction, indicating that the 2D vdW heterostructures are thermally robust. A snapshot of the final atomic configuration of 2D LiAlTe$_2$/GeTe vdW heterostructures from the MD simulation is shown in Figure S2. The corresponding total energy fluctuation with time during the simulation shows no evident structural instability.

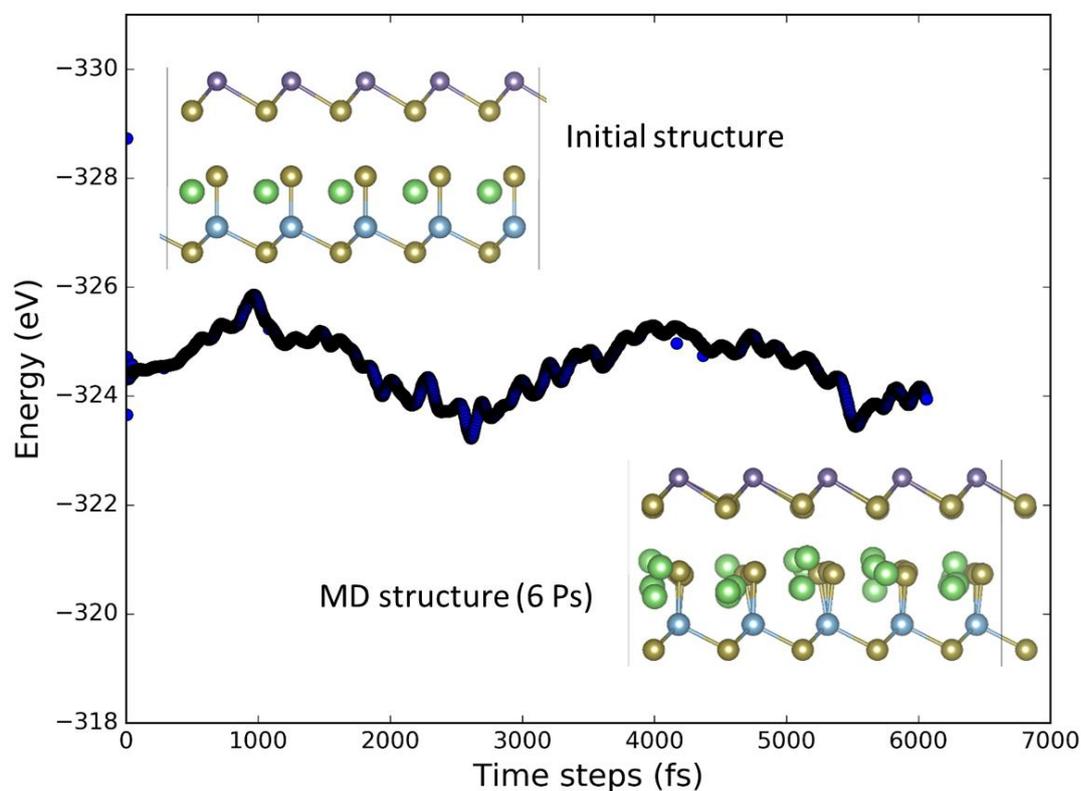

**Figure S2.** Fluctuations of total energy as evolution of simulation time and the snapshots of the initial and final atomic configurations (side view) after the first-principles molecular dynamics (MD) simulations (6 ps) with time step of 1 fs at the temperature of 300 K.

23